\documentclass[usenatbib]{mn2e}
\usepackage{amsmath}
\usepackage{pifont}
\usepackage{amssymb}
\usepackage{graphicx}
\usepackage{verbatim}
\usepackage{natbib}

\def\gsim{\;\lower4pt\hbox{${\buildrel\displaystyle >\over\sim}$}\;}
\def\lsim{\;\lower4pt\hbox{${\buildrel\displaystyle <\over\sim}$}\;}
\def\grls{\;\lower4pt\hbox{${\buildrel\displaystyle >\over <}$}\;}

\title[X-ray Emissions of Shocked Dynamic PNe]
{Diffuse X-ray Emissions from Dynamic Planetary Nebulae}

\author[Y.-Q. Lou \& X. Zhai]{Yu-Qing Lou$^{1,2,3,}$\thanks{Email:
louyq@tsinghua.edu.cn; lou@oddjob.uchicago.edu}
and Xiang Zhai$^{1,4}$\thanks{zxzhaixiang@gmail.com}\\
$^{1}$ Department of Physics and Tsinghua Centre for
Astrophysics (THCA), Tsinghua University, Beijing, 100084, China;\\
$^{2}$ Department of Astronomy and Astrophysics, the University of
Chicago, 5640 S. Ellis Avenue, Chicago, IL 60637, USA;\\
$^{3}$ National Astronomical Observatories, Chinese Academy of
Sciences, A20, Datun Road, Beijing, 100021, China;\\
$^{4}$ Department of Physics, University of California at San Diego,
9500 Gilman Drive, La Jolla, CA 92037-0350, USA}
\begin{document}
\bibliographystyle{agms}
\date{Accepted 2010 June 2.\ \ Received 2010 May 28;\ \
in original form 2010 May 4 } \maketitle

\begin{abstract}
We present theoretical results of a piecewise isothermal shock
wind model devised for predicting the luminosity and surface
brightness profile of diffuse X-ray emissions primarily from the
inner shocked downstream wind zone of a planetary nebula (PN)
surrounded by self-similar shocked dense shell and outer slow AGB
wind envelope involving self-gravity and compare/fit our
computational model results with available observations of a few
grossly spherical X-ray emitting PNe. Matching shocked piecewise
isothermal self-similar void (ISSV) solutions with self-gravity of
Lou \& Zhai (LZ) for the outer zone and a stationary isothermal
fast tenuous wind with a reverse shock for the inner zone across
an expanding contact discontinuity, we can consistently construct
dynamic evolution models of PNe with diffuse X-ray emissions. On
the basis of such a chosen dynamic wind interaction model, both
X-ray luminosity and radial X-ray brightness profile are
determined by three key parameters, namely the so-called X-ray
parameter $X$, two radii $R_{rs}$ and $R_c$ of the reverse shock
and the contact discontinuity. We find that morphologies of X-ray
emissions would appear in the forms of either a central luminous
sphere or a bright ring embedded within optically bright shells.
In contrast to previous adiabatic models, the X-ray brightness
peaks around the reverse shock, instead of the contact
discontinuity surface just inside the outer shocked dense shell.
Diffuse X-ray emissions of a few observed PNe appear to support
this wind-wind dynamic interaction scenario.
\end{abstract}

\begin{keywords}
hydrodynamics --- ISM: clouds --- planetary nebulae
--- radiation --- shock waves --- X-rays: ISM
\end{keywords}

\section{Introduction}

Planetary nebulae (PNe) are produced at the late stage of stellar
evolution, when a star transits from the asymptotic giant branch
(AGB) phase where a star blows slow AGB dense wind to the later
phase showing a central compact star (i.e. hot white dwarf
gradually exposed).
As the later tenuous fast wind chases and collides with an earlier
circumstellar slow massive AGB wind envelope, a forward shock
emerges and runs outwards into the slow AGB wind zone, while a
reverse shock travels radially inwards relative to the inner fast
wind; a contact discontinuity in between separates the forward and
reverse shocks. Such dynamic shock interaction of wind-wind
collision rearranges the distribution of AGB wind envelope and
eventually result in a quasi-spherical PN \citep*[e.g.,][-- referred
to as LZ hearafter]{Kwoketal1978, VolkKwok1985, Chevalier1997,
BianLou2005, Yuetal2006, LZ}. Ignoring the gravity effect
completely, Chevalier (1997) presented an isothermal dynamic model
for PNe and constructed spherically symmetric global hydrodynamic
solutions to describe the expansion of outer shocked shell with an
inner contact discontinuity of fast wind moving at a constant speed
in the inner zone. \citeauthor{LZ} advanced a piecewise isothermal
model framework including the gas self-gravity and showed various
plausible self-similar dynamic behaviours of outer PN envelopes. The
dynamic collision interaction of the inner fast tenuous wind with
the outer slow AGB wind envelope is well taken into account by
including forward and reverse shocks and contact discontinuity in
\citeauthor{LZ}. As an illustrating example of PN application,
isothermal self-similar void (ISSV) solutions of \citeauthor{LZ}
appear sensible in fitting the available data of the PN NGC 7662.

Diffuse X-ray emissions due to spectral line cooling of highly
ionized metals (e.g. Kastner et al. 2000; Yu et al. 2009) and
probably partial thermal bremsstrahlung from hot electrons reveal
important inner morphological features of PNe, i.e. the region
enclosed within the dense shell. A powerful reverse shock can heat
up the inner tenuous fast wind plasma on the downstream side
within an estimated temperature range of $\sim 10^6-10^8$ K (e.g.
\citealt{VolkKwok1985}; \citealt{Akashietal2006}; \citealt{Leahyetal2000};
\citealt{StuteSahai2006}) and create a reverse shock heated `hot bubble' largely
confined by the outer dense shell at an expanding contact
discontinuity. For this overall physical scenario, properties of
diffuse X-ray emissions are mainly determined by the temperature,
density, velocity and mass loss rate of the shocked inner wind. As
a result, it offers unique insight for the shaping processes of
PNe and their central hot bubbles.

In the past decade, advanced instruments in space allow us to
detect diffuse X-ray emissions from PNe. Together with
complementary observations for dense shells of PNe in optical and
infrared bands, we are now able to test the wind-wind dynamic
shock interaction scenario on more specific terms. X-ray emissions
from three PNe were detected with \emph{ROSAT} spacecraft (e.g.
Stute \& Sahai 2006), and from nine PNe with \emph{Chandra} and
\emph{XMM-Newton} X-ray Observatories (e.g. Kastner et al. 2000,
2001, 2003; Chu et al. 2001; Guerrero et al. 2002, 2005; Sahai et
al. 2003; Montez et al. 2005; Gruendl et al. 2006). With
sufficiently high spatial and energy resolutions as well as
detector sensitivity, \emph{Chandra} and \emph{XMM-Newton}
spacecraft observations allow us to study the extended spatially
resolved X-ray structure of PNe, which provide an unprecedented
opportunity to examine the wind-wind dynamic interaction scenario.

Akashi et al. (2006) qualitatively calculated the morphology of
X-ray emissions from PNe
based on the adiabatic self-similar 
colliding wind solutions for shocked fast wind flows of
\citet{ChevalierImamura1983}without self-gravity and showed that the
X-ray emission appears in the form of a narrow ring inside the
optically bright part of a PN. This was suggested to be the case of
NGC 40 by Montez et al. (2005).
Akashi et al. (2007) investigated X-ray emissions from PNe by
numerical simulation of adiabatic winds without gravity. They
found that in order to explain the observed PNe X-ray properties,
a rapidly decreasing mass loss rate is required. The adiabatic
colliding wind models of Chevalier \& Imamura (1983) and Akashi et
al. (2007) both indicate that X-ray emissions become brightest
very near the dense shell. Therefore, PNe with central luminous
X-ray emission profiles cannot be accounted for by their adiabatic
colliding wind models. Chevalier \& Imamura (1983) also included
an isothermal dense shell in their model, in addition to the
innermost and outermost stationary isothermal winds.

Recent observations reveal that diffuse X-ray emissions from PNe
do generally lie within the interior of the optically bright
shells of relatively dense ionized gas, such as PN BD
+30$^{\circ}$3639 (e.g. Kastner et al. 2000; Yu et al. 2009), NGC
40 (e.g. Montez et al. 2005) and NGC 2392 (e.g. Guerrero et al.
2005).
From PN BD +30$^{\circ}$3639, ROSAT spacecraft instrument
discovered a Gaussian-shape diffuse X-ray emission source (e.g.
Leahy et al. 2000). With a high spatial resolution of $0''.49$,
Chandra spacecraft well resolved this compact X-ray emission
source and found a quasi-spherical central bright diffuse X-ray
emission blob embedded inside the optical bright shell, with a
radius $\sim 2''$. The X-ray photon flux from the centre is $\sim
10$ times more intense than that from $\sim 2''$ (e.g. Kastner et
al 2000). Moreover, Chandra also found the X-ray bright centre was
slightly concentrated towards the east side. Different from PN BD
+30$^{\circ}$3639, PN NGC 40 shows a distinct annulus X-ray
emission e.g. Montez et al. 2005).
For those PNe of bipolar
morphologies that do not clearly present shell-like structures,
their diffuse X-ray emissions are also from within the optically
bright shells, such as NGC 7027 (e.g. Kastner et al. 2001), NGC
7009 (e.g. Guerrero et al. 2002) and NGC 6543 (e.g. Chu et al.
2001), NGC 7026 (e.g. Gruendl et al. 2006) and Menzel 3 (e.g.
Kastner et al. 2003). These X-ray observations bear strong
implications that wind-wind dynamic interactions or jet-wind
interactions are most likely responsible for the inner diffuse
X-ray emissions from PNe (e.g. Kastner 2007).

This paper establishes a quantitative connection of self-similar
piecewise isothermal dynamic evolution model of PNe with diffuse
X-ray emissions, especially the radial distribution of surface
brightness. On the basis of \citeauthor{LZ} global ISSV model for
the self-similar evolution of PNe, we show that three key model
parameters determine the morphology for diffuse X-ray emissions from
PNe and specifically examine observations of PN NGC 40.
In our piecewise isothermal model framework, either ring-like or
central bright X-ray morphologies are possible because X-ray
emission peaks near the reverse shock instead of the contact
discontinuity.
Our results lend support to the wind-wind dynamic interaction
model for PNe in general. We also compute the luminosity, surface
brightness of X-ray emission from the PN 7662 and emphasize that
this PN should be a highly desirable target source for
\emph{Chandra} further X-ray observations in the near future.

\section{A spherical wind-wind dynamic interaction
model for Planetary Nebulae with self-similar shells}

In this section, we first describe our global piecewise isothermal
dynamic evolution model for quasi-spherical PNe. At the late stage
of stellar evolution, a star enters the
AGB phase. During this phase, a star suffers from a rather high rate
of mass loss $\sim 10^{-6}$M$_{\odot }\/$ yr$^{-1}$ and thus creates
a fairly dense AGB wind shell expanding at a slow radial speed of
$\sim 10$ km s$^{-1}$ \citep[e.g.][\citeauthor{LZ}]{Kwoketal1978,
ChevalierImamura1983, Chevalier1997}. The dense gas shell keeps
expanding until the stellar hydrogen envelope has been almost
depleted and a compact hot white dwarf gradually exposed at the
centre.
Meanwhile, a hot tenuous fast wind somehow emerges from the central
white dwarf as indicated by observations. This tenuous fast wind
blows at a high speed of $\gsim 10^3$ km s$^{-1}$ with a mass loss
rate of $\sim 10^{-10}-10^{-7}$ M$_{\odot }\/$ yr$^{-1}$ (e.g.
Cerruti-Sola \& Perinotto 1985). This fast inner wind will not
diffuse through the outer AGB slow wind, but smash onto the outer
dense AGB wind shell (e.g. Kwok et al. 1978) and then give rise to a
forward shock propagating outwards into the slow AGB wind and a
reverse shock propagating inwards in the fast wind comoving
framework of reference (e.g. \citet{ChevalierImamura1983}; Akashi et
al. 2006). A contact discontinuity naturally forms when the gas
pressures of fast wind and of AGB wind reach a dynamic balance with
both density and temperature being different on the two adjacent
sides. These two shocks, especially the forward shock, rearranges
the distribution of gas flow and finally results in a dense shell
surrounding the central compact star blowing an inner fast wind
(e.g. \citealt{Kwok2000}; \citeauthor{LZ}).

A quasi-spherical PN may be roughly divided into four regions (see
figure 1 of Akashi et al. 2006). The fast wind occupies the inner
region from the central compact white dwarf to the reverse shock
front at radius $R_{rs}$. When propagating inwards relative to the
fast wind, the reverse shock heats up and slows down the inner wind
and creates a downstream wind region from the reverse shock front to
the expanding contact discontinuity at radius $R_c$. These two
regions form the inner zone and are characterized by the fast wind
mass loss rate, wind velocities, densities, temperatures and reverse
shock speed. The shocked dense shell region stays between the
contact discontinuity and the forward shock at radius $R_{s}$. The
fourth region, or the outer AGB slow wind envelope, expands outside
the forward shock front at a low speed of $\sim 10$ km s$^{-1}$. As
in \citeauthor{LZ}, we adopt spherically symmetric nonlinear
hydrodynamic partial differential equations (PDEs) in spherical
polar coordinates $(r,\ \theta,\ \varphi)$ to model the piecewise
isothermal gas dynamics of these four wind regions of a
quasi-spherically symmetric PN.

\subsection{Inner Shocked Tenuous Wind Zone\\ \qquad\
Featuring a Reverse Shock}

As already noted, a reverse shock emerges and propagates in the
inner fast wind zone due to the collisional interaction between
the later fast and earlier slow winds. The mass and radial
momentum conservations across an isothermal shock front in the
shock comoving framework are simply
\begin{multline}\label{general_shock}
\rho_d(u_d-u_s)=\rho_u(u_u-u_s)\ ,\\
\!\!\!\!\!\!\!\!\!\!\!\!\!\!\!\!
a_d^2\rho_d+\rho_du_d(u_d-u_s)=a_u^2\rho_u+\rho_uu_u(u_u-u_s)\ ,
 \ \ \qquad
\end{multline}
where $a$ is the isothermal sound speed, $\rho$ is the mass density
and $u$ is the radial flow velocity, and the two subscripts $_d$ and
$_u$ denote the downstream and upstream sides of an isothermal shock
and the subscript $s$ indicate association with the shock
\citep[e.g.][\citeauthor{LZ}]{CourantFriedrichs1976, Spitzer1978,
ShenLou2004, BianLou2005, Yuetal2006}. The temperature ratio across
such a shock front is another parameter to be specified
\citep[see][]{ShenLou2004}. The reverse shock travels inwards
relative to the inner wind, so the inner fast wind region is the
upstream side of the shock and the reverse shock heated wind
occupies the downstream side until the contact discontinuity at the
expanding radius $R_c$. In the laboratory framework of reference, a
reverse shock may move inwards or outwards, or appear stationary in
space depending on various specific situations
\citep[e.g.,][]{wispI, wispII}.

We presume a constant mass loss rate $\dot{M}_{fw}$ for an
isothermal inner fast wind and define $a_{w,d(u)}$, $T_{w,d(u)}$,
$\rho_{w,d(u)}$ and $v_{w,d(u)}$ as the isothermal sound speed,
constant temperature, mass density and constant wind velocity on
the downstream (upstream) side of the reverse shock. The sound
speed ratio $\tau_w\equiv a_{w,d}/a_{w,u}=(T_{w,d}/T_{w,u})^{1/2}$
across the shock front characterizes the downstream plasma heating
of the reverse shock. Isothermal shock conditions
(\ref{general_shock}) then give
\begin{multline}\label{reverse_shock}
v_{w,u}-v_{rs}=({1}/{2})[v_{w,d}-v_{rs}
+{a_{w,d}^2}/{(v_{w,d}-v_{rs})}]\\
+\frac{1}{2}\left\lbrace\left[\frac{(v_{w,d}-v_{rs})^2
-a_{w,d}^2}{(v_{w,d}-v_{rs})}\right]^2
+4a_{w,d}^2\frac{\tau_w^2-1}{\tau_w^2}\right\rbrace^{1/2},\\
v_{w,d}-v_{rs}=({1}/{2})[v_{w,u}-v_{rs}
+{a_{w,u}^2}/{(v_{w,u}-v_{rs})}]\\
-\frac{1}{2}\left\lbrace\left[\frac{(v_{w,u}-v_{rs})^2
-a_{w,u}^2}{(v_{w,u}-v_{rs})}\right]^2
+4a_{w,u}^2(1-\tau_w^2)\right\rbrace^{1/2},\\
\rho_{w,d}(R_{rs},t)=\rho_{w,u}(R_{rs},t){(v_{w,u}-v_{rs})}/{(v_{w,d}
-v_{rs})}\ ,
\end{multline}
where $v_{rs}$ is the reverse shock velocity in the laboratory
reference framework (\citeauthor{LZ}).
Calculations show that positive, negative and zero $v_{rs}$ are
all physically allowed. For $v_{rs}<0$, the inner fast wind region
shrinks radially inwards until
a certain epoch. In other words, the reverse shock heats up the
downstream wind plasma and leads to a hot bubble inside the
contact discontinuity surface at an expanding radius $R_c$.

The mass density profile within the central fast wind zone is
simply
\begin{equation}\label{fastwind_density}
 \rho_{w,u}(r,\ t)={\dot{M}_{fw}}/({4\pi v_{w,u}r^2})
\end{equation}
and that in the reverse-shocked downstream wind zone is
\begin{equation}\label{shocked_fastwind_density}
 \rho_{w,d}(r,t)=\frac{1-v_{rs}/v_{w,u}}{(v_{w,d}-v_{rs})}
 \frac{\dot{M}_{fw}}{4\pi r^2}\equiv\frac{\beta}{r^2}
 \approx\frac{\dot{M}_{fw}r^{-2}}{4\pi (v_{w,d}-v_{rs})}
\end{equation}
which remains steady, where the constant $\beta$ parameter is
explicitly defined.
This last approximation highlights that the downstream density of
shocked inner wind does not sensitively depend on the inner fast
wind speed $v_{w,u}$ because we typically have a small speed ratio
of $v_{rs}/v_{w,u}\sim 10^{-2}$.
Due to the reverse shock, the downstream mass loss rate just
inside the contact discontinuity surface is given by
\begin{equation}
\dot{M'}=4\pi R_c^2v_{w,d}\rho_{w,d}(R_c,\
t)=\frac{(1-v_{rs}/v_{w,u})}{(1-v_{rs}/v_{w,d})}\dot{M}_{fw}\ ,
\end{equation}
which is equal to the central mass loss rate $\dot{M}_{fw}$ only
when the reverse shock remains stationary in the laboratory
reference framework. The difference between $\dot{M}_{fw}$ and
$\dot{M'}$ is the mass loss rate at the reverse shock front due to
the reverse shock movement.

X-ray observations in the past several years appear to indicate
that the temperatures of inner shocked fast wind, where X-ray
photons are released, are a factor of $\sim 2-10$ lower than that
predicted by simple energy conservation arguments (e.g. Soker \&
Kastner 2003). In our formulation, we use a chosen sound speed
ratio $\tau_w\equiv a_{w,d}/a_{w,u}=(T_{w,d}/T_{w,u})^{1/2}$ to
bridge the two constant temperatures of upstream and downstream
sides across a moving reverse shock.

\subsection{The Dense Shell Region and the
\\ \qquad\ Outer AGB Slow Wind Envelope}

The shocked dense shell region and the outer AGB slow wind envelope
separated by an outgoing forward shock are described by the ISSV
model of \citeauthor{LZ}. The piecewise isothermal hydrodynamic
equations with spherical symmetry are the mass conservation
\begin{equation}\label{MCE}
  \frac{\partial M}{\partial t}+u\frac{\partial M}{\partial r}=0\ ,
  \qquad\qquad\qquad
  \frac{\partial M}{\partial r}=4\pi r^2\rho\ ,
\end{equation}
or equivalently,
\begin{equation}
  \frac{\partial\rho}{\partial t}
  +\frac{1}{r^2}\frac{\partial}{\partial r}(r^2\rho
  u)=0\, \label{CE}
\end{equation}
and the radial momentum equation
\begin{equation}
  \frac{\partial u}{\partial t}+u\frac{\partial u}{\partial r}
  =-\frac{a^2}{\rho}\frac{\partial\rho}{\partial
  r}-\frac{GM}{r^2}\ ,\label{ME}
\end{equation}
where $u$ is the radial bulk flow speed; $M(r,\ t)$ is the
enclosed mass within radius $r$ at time $t$; $\rho(r,\ t)$ is the
mass density; $a\equiv(p/\rho)^{1/2}=(k_{\rm B}T/\bar m)^{1/2}$ is
the isothermal sound speed; $T$ is the constant gas temperature;
$\bar m$ is the mean particle mass; $p$ is the gas pressure,
$k_{\rm B}$ is Boltzmann's constant and $G=6.67\times 10^{-8}$
cm$^3/$(g s$^2$) is the gravitational constant.

By the known isothermal self-similar transformation
\begin{multline}\label{ST}
   \ \ x={r}/{(at)}\ ,\qquad \qquad \qquad \
   \rho(r,\ t)=\alpha(x)/(4\pi G t^2)\ ,\\
\!\!\!\!\!\!\!\!
   M(r,\ t)=a^3tm(x)/G\ ,\qquad\quad  u(r,\ t)=av(x)\ ,
   \ \qquad
\end{multline}
equation (\ref{MCE}) leads to
\begin{equation}
m(x)=x^2\alpha(x-v)\ ;\label{enclosedmass}
\end{equation}
and then PDEs (\ref{CE}) and (\ref{ME}) reduce to two coupled
nonlinear ordinary different equations (ODEs)
\begin{equation}
\left[(x-v)^2-1\right]\frac{dv}{dx}=\left[\alpha(x-v)
-\frac{2}{x}\right](x-v)\ ,\label{ODE1}
\end{equation}
\begin{equation}
\left[(x-v)^2-1\right]\frac{1}{\alpha}\frac{d\alpha}{dx}
=\left[\alpha-\frac{2}{x}(x-v)\right](x-v)\ ,\label{ODE2}
\end{equation}
where $\alpha(x)$, $m(x)$, $v(x)$ are the dimensionless reduced
variables corresponding to mass density $\rho(r,\ t)$, enclosed mass
$M(r,\ t)$ and radial flow speed $u(r,\ t)$, respectively
\citep{Shu1977, Hunter1977, WhitworthSummers1985, TsaiHsu1995,
Shuetal2002, LouShen2004, ShenLou2004, BianLou2005, LZ}. Given
proper analytic asymptotic solution conditions at large and small
$x$ as well as in the neighborhood of the sonic critical point,
these two coupled nonlinear ODEs (\ref{ODE1}) and (\ref{ODE2}) can
be solved in a straightforward manner. Each dimensionless ISSV
solution of these two ODEs gives a behaviour of the dense shell and
outer AGB wind envelope outside $R_c$.

The forward shock travels outwards in a self-similar manner and
divides the shell into the shocked dense shell region and the outer
slow AGB wind envelope. We define $a_{d(u)}$, $T_{d(u)}$,
$\alpha_{d(u)}$ and $v_{d(u)}$ as the sound speed, temperature,
reduced mass density and reduced gas radial flow velocity on the
downstream (upstream) side of the forward shock and the sound speed
ratio $\tau\equiv a_d/a_u=(T_d/T_u)^{1/2}$
\citep[\citeauthor{LZ}]{ShenLou2004, BianLou2005, Yuetal2006}. In
this case, the dense shell region is the downstream side and the
outer slow AGB wind envelope is the upstream side of the outgoing
forward shock front. We define dimensionless variables $x_{sd}\equiv
V_s/a_d$ and $x_{su}\equiv V_s/a_u$ where $V_s=r_s/t$ is the
radially outgoing velocity of the forward shock in the laboratory
reference framework. Then isothermal shock conditions
(\ref{general_shock}) bear the self-similar form of
\begin{multline}\label{ss_shock}
  \alpha_d/\alpha_u=(v_u-x_{su})/[\tau(v_d-x_{sd})]\ ,\\
\!\!\!\!\!\!\!\!\!
  (v_d-x_{sd})/(v_u-x_{su})-\tau =(\tau v_d-v_u)(v_d-x_{sd})\
  \quad
\end{multline}
(Shen \& Lou 2004). These isothermal shock jump conditions connect
the AGB outer wind envelope with the dense shell region across the
forward shock front.

The analytic asymptotic solution of coupled nonlinear ODEs
(\ref{ODE1}) and (\ref{ODE2}) at large $x$ is known as
\begin{multline}\label{largeasymp}
    v=V+\frac{2-A}{x}+\frac{V}{x^2}
    +\frac{(A/6-1)(A-2)+2V^2/3}{x^3}+\cdots\ ,\\
    \alpha=\frac{A}{x^2}+\frac{A(2-A)}{2x^4}
    +\frac{(4-A)VA}{3x^5}+\cdots\ ,\qquad\qquad\ \
\end{multline}
where $V$ and $A$ are the velocity and mass parameters
(\citealt{LouShen2004}; \citeauthor{LZ}).
In our self-similar piecewise isothermal model framework, these
two parameters characterize the asymptotic massive AGB wind
velocity as
\begin{equation}\label{v_AGB}
v_{\rm AGB}=Va_u
\end{equation}
at very large $x$ and a constant AGB wind mass loss rate as
\begin{equation}\label{AGB_massloss}
\dot{M}_{\rm AGB}=VAa_u^3/G\ .
\end{equation}
These asymptotic relations are used to constrain the range of
model parameters for PNe.

\subsection{Radial Expansion of the Spherical
\\ \ \qquad Contact Discontinuity Interface}

The contact discontinuity surface at $R_c$ expands at a constant
speed and separates the inner and outer zones. \citeauthor{LZ}
estimated that the gravity of inner tenuous wind zone may be
negligible during the dynamic self-similar evolution of the dense
shell region and outer AGB envelope in outer zone. By letting
$m(x_0)=0$ in equation (\ref{enclosedmass}), we then obtain the
contact discontinuity surface radius $R_c=a_dx_0t$
and its expansion velocity $v_c=a_dx_0$.
By doing this, we
ignore the self-gravity from the inner shocked wind region within
the contact discontinuity radius $R_c$, but still retain the
self-gravity of the dense shell and the AGB slow wind envelope
outside the contact discontinuity radius $R_c$.
A contact discontinuity requires that the inner downstream wind
velocity $v_{w,d}$ and the outer dense shell gas velocity at $R_c$
be equal to the contact discontinuity expansion velocity, namely
\begin{equation}\label{CD_requirement}
v_{w,d}=a_dx_0=a_dv(x_0)\ ,
\end{equation}
which is satisfied automatically when $m(x_0)=0$ occurs. Here,
$x_0$ is where an ISSV solution for the dense shell region and the
outer envelope starts. Given a proper set of $\{x_0,\
\alpha_0\equiv\alpha(x_0),\ x_{ds}\}$, one can readily integrate
coupled nonlinear ODEs (\ref{ODE1}) and (\ref{ODE2}) from $x_0$ to
larger $x\ge x_0$ (i.e. from left to right), apply self-similar
shock jump conditions (\ref{ss_shock}) at $x_{ds}$ to cross an
outgoing forward shock front once $a_d$ and $a_u$ are known and
then continue further from $x_{us}$ to sufficiently large $x$ to
determine velocity and mass parameters $V$ and $A$ with desired
accuracies.
The flow zone between the contact discontinuity and the forward
shock is the dense downstream side and the outer AGB slow wind is
the upstream side. The forward shock jump condition connects
$x_{ds}$ and $x_{us}$.
These solutions have a negligible mass within $x_0$ and are referred
to as ISSV solutions by \citeauthor{LZ}. In total, eight parameters
are involved for constructing a model ISSV solution
$\{x_0, \alpha_0, a_d, x_{ds}, a_u, x_{us}, V, A\}$,
with five of them being actually independent.

At radius $R_c$ of contact discontinuity, the inner downstream
wind plasma pressure is
\begin{equation}
P_{w,d}(R_c,\ t)=\frac{k_BT_{w,d}}{\bar
m_{w,d}}\frac{\beta}{R_c^2}\propto\frac{T_{w,d}}{t^2}\ ,
\end{equation}
where $\bar m_{w,d}$ is the mean particle mass of the inner
downstream wind zone. With self-similar transformation (\ref{ST}),
we have the shocked dense shell gas pressure at the contact
discontinuity radius $R_c$ as
\begin{equation}
P_d(R_c,\ t)=\frac{k_BT_d}{\bar m_d}\frac{\alpha_0}
 {4\pi Gt^2}\propto\frac{T_d}{t^2}\ ,
\end{equation}
where $\bar m_d$ is the mean particle mass of the shocked dense
shell zone. To physically maintain an expanding contact
discontinuity at constant speed, the parameters $\dot{M}_{fw}$,
$v_{rs}$, $v_{w,d}$, $v_{w,u}$,
$T_{w,d}$, $\alpha_0$ and $T_d$ are coupled by the pressure
balance condition $P_{w,d}(R_c,t)=P_d(R_c,t)$ across the contact
discontinuity, namely
\begin{equation}\label{CD_balance}
\frac{T_d}{\bar m_d}\frac{\alpha_0}{G}=\frac{T_{w,d}}{\bar
m_{w,d}}\frac{4\pi\beta}{v_{w,d}^2}=\frac{T_{w,d}}{\bar m_{w,d}}
\frac{(1-v_{rs}/v_{w,u})}{(1-v_{rs}/v_{w,d})}\frac{\dot{M}_{fw}}{
v_{w,d}^3}\ .
\end{equation}
In general, the two mean particle masses $\bar m_d$ and $\bar
m_{w,d}$ are allowed to be different depending on specific
situations.

\subsection{Comparison with Previous Models of PNe}

\citet{ChevalierImamura1983} and \citeauthor{LZ} both invoke a
self-similar hydrodynamic wind-interaction phase to explore the
global dynamic shock evolution of PNe.
In both models, the inner and outer wind zones are characterized
by different yet constant mass loss rates.


More specifically, in the model of Chevalier \& Imamura (1983),
the innermost fast wind zone and the outermost AGB slow wind
envelope are both isothermal and expand at constant speeds with
both mass densities scaling as $\propto r^{-2}$. In between these
two wind regions, both shocked downstream wind zone and dense
shell are solved separately by adopting a type of adiabatic
self-similar transformation for $\gamma=5/3$ hydrodynamic PDEs
without gas self-gravity (Parker 1961). These two self-similar
dynamic zones are joined by the pressure balance across an
outgoing contact discontinuity. Two different Mach numbers are
specified respectively for the reverse shock separating inner
$r^{-2}$ fast wind and self-similar shocked downstream wind zone
(hot bubble), and for the forward shock separating self-similar
dense shell and outer $r^{-2}$ slow AGB wind envelope.

In comparison, \citeauthor{LZ} model also presumes a $r^{-2}$
density profile for a steady piecewise isothermal fast wind zone in
the central region around the remnant white dwarf. Differently, the
dense shell and the outer AGB slow wind envelope are described by a
self-similar dynamic solution with self-gravity and an isothermal
shock jump (i.e. type $\mathcal{Z}$ self-similar solutions of
\citeauthor{LZ}). The self-similar process here differs from that of
Chevalier \& Imamura (1983) in (a) the inclusion of self-gravity;
(b) the piecewise isothermal wind (both the dense shell and the
outer AGB slow wind envelope are isothermal but with different
temperatures);
(c) a different self-similar transformation due to the presence of
self-gravity and (d) necessary shock parameters (forward shock is
described by its velocity and sound speed ratio or temperature jump
parameter $\tau$). In the inner shocked downstream wind zone,
\citeauthor{LZ} model has constant temperature and wind speed. For a
constant mass loss rate $\dot{M}_{fw}=\rho_{w,d}v_{w,d}4\pi
r^2(1-v_{rs}/v_{w,d})/(1-v_{rs}/v_{w,u})$,
both mass density and thermal gas pressure scale as $r^{-2}$.

In the inner shocked downstream wind zone and the dense shell,
\citeauthor{LZ} piecewise isothermal model behaves differently from
adiabatic self-similar solutions of Chevalier \& Imamura (1983). For
Chevalier \& Imamura solution of colliding winds, mass density and
pressure in both downstream zones (i.e., the inner shocked
downstream wind and the dense shell) are lowest near the shock
fronts and highest near the contact discontinuity where strong
shocks were initiated. Under weak shock condition, density behaves
similarly but pressure is no longer monotonic, which may lead to
Rayleigh-Taylor instability \citep[e.g.][]{ChevalierImamura1983}.
Several isothermal cases with $\gamma=1$ for the dense shell are
also calculated by Chevalier \& Imamura (1983) to find that both
density and pressure peak at the forward shock front, instead of at
the contact discontinuity, which is qualitatively similar to
\citeauthor{LZ} solutions
for the dense shell. Another difference is the behaviour of AGB
envelopes. In \citeauthor{LZ} model, both velocity and mass loss
rate of AGB envelopes approach constant at large radii.

Based on Chevalier \& Imamura (1983) adiabatic solution for inner
shocked fast wind, Akashi et al. (2006) find the morphology of the
X-ray emission is in the form of a narrow ring just inside the
optical bright part of a PN (i.e. dense shell). However, simulation
shows that the velocity of reverse shock and contact discontinuity
cannot stably remain constant (Akashi et al. 2007). This might be
related to the Rayleigh-Taylor instability for some adiabatic
solutions of Chevalier \& Imamura (1983). Akashi et al. (2007) also
find that in order to get more realistic simulation results, a
rapidly decreasing mass loss rate of inner fast wind is required. In
this case, the density should be no longer increasing from the
reverse shock to the contact discontinuity. For Chevalier \& Imamura
isothermal solutions and for \citeauthor{LZ} model, mass densities
increase monotonically from the contact discontinuity to the shock
front. A physical consequence is that near the reverse shock gas is
brighter in X-ray emissions than near the contact discontinuity.

For a PN, both forward and reverse shocks are radiative
\citep[e.g.][]{ChevalierImamura1983}. The central hot white dwarf
can be a strong source of photoionizing radiation and may completely
photoionize the outer dense shell (e.g. Chevalier 1997;
\citeauthor{LZ}). Therefore, in reality, the hot bubble and dense
shell should involve gas dynamics between adiabatic and isothermal
processes (say, a polytropic description as an approximation).

\section{Diffuse X-ray Emissions from PNe}

The reverse shock initiated by the collision of an inner fast
tenuous wind with ambient dense gas materials previously blown out
during the AGB slow wind phase heats up the inner downstream wind
region, leading to
spectral line emissions from highly ionized metals such as C, O, Ne
in X-ray bands (e.g. Volk \& Kwok 1985; Akashi et al. 2006; Leahy et
al. 2000; Stute \& Sahai 2006; Yu et al. 2009). In contrast,
temperatures of the shocked dense shell and of the outer AGB wind
envelope are only in the order of $\sim 10^4$ K (e.g.
\citealp{Guerreroetal2004}; \citeauthor{LZ}), and that of the
central hot white dwarf is $\lsim 2\times 10^5$ K (e.g. Werner et
al. 2008). Consequently, X-ray emissions from PNe would be mainly
produced in the inner shocked downstream wind region with a higher
temperature.
In this scenario, X-ray emissions from PNe projected in the plane
of sky should largely lie within the expanding contact
discontinuity of radius $R_c$.

\begin{figure}
\centering
\includegraphics[width=0.50\textwidth]{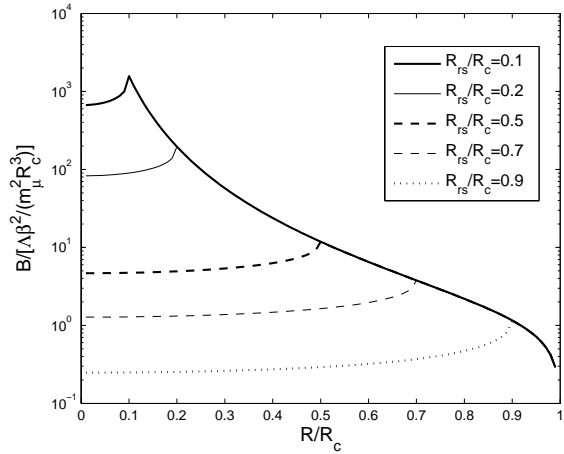}
\caption{Shown here are variations of the projected dimensionless
X-ray brightness radial profile $B/[\Lambda\beta^2/
(m_{\mu}^2R_c^3)]$ within the contact discontinuity radius $R_c$
for PNe with different radius ratios $R_{rs}/R_c$. The projected
radius $R$ is also normalized by the contact discontinuity radius
$R_c$. The range of brightness $B$ variation spans three order of
magnitudes as $R_{rs}/R_c$ varies from 0.1 to 0.9. Different
epochs of evolution might be invoked to explain why some PNe are
observed to be strong X-ray sources while other sources of PNe are
not. }\label{fig:XrayBrightness}
\end{figure}

\begin{figure}
\centering
\includegraphics[width=0.50\textwidth]{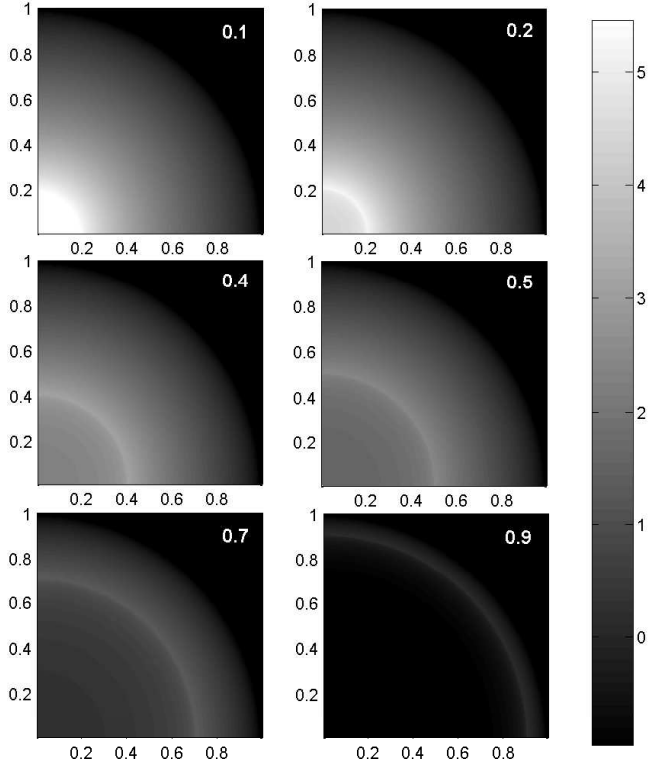}
\caption{Gray-scale plots of the logarithm of dimensionless X-ray
surface brightness (see Fig. \ref{fig:XrayBrightness}) with
different radius ratios $0.1\leq R_{rs}/R_c\leq 0.9$ marked at the
upper-right corner of each panel (see Fig.
\ref{fig:XrayBrightness}). The
abscissa and the
ordinate represent normalized projected radius $R/R_c$ by the
contact discontinuity radius $R_c$. As the radius ratio
$R_{rs}/R_c$ may vary with time, the corresponding variations in
brightness profile is thus expected.
}\label{fig:XrayMap}
\end{figure}

According to \citeauthor{LZ}, the typical electron number density in
the dense shell zone is estimated as $n_e\sim 10^3(1000$yr$/t)^2$
cm$^{-3}$, where $t$ is the typical age of a PN, and the thickness
of a dense shell zone is estimated as $\sim 10^{18}(t/1000$yr) cm.
Then the mean free path of a photon in the dense shell zone would be
$l=1/(n_e\sigma)\sim 10^{20}(t/1000$yr$)^2$ cm, where
$\sigma=6.65\times 10^{-25}$ cm$^2$ is the electron cross section
for Thomson scattering. Therefore the dense gas shell should be
optically thin for emissions of X-ray photons during the evolution
of a PN. We denote the X-ray emissivity of the shock heated
downstream inner wind plasma by $\Lambda(T)$ such that $\Lambda
n_p^2$ is the energy emitted per unit time per unit volume, where
$n_p$ is the proton number density (e.g. Sarazin 1986). At the
temperature range of $T\lsim 3\times 10^7$ K when spectral line
emissions from highly ionized metals are important, a simple
approximation is thus $\Lambda(T)\approx 6.2\times 10^{19}T^{-0.6}$
ergs cm$^{-3}$ sec$^{-1}$ (e.g. Sarazin 1986). The X-ray luminosity
$L_X$ of a PN can then be calculated by integrating the emissivity
over the entire shocked downstream inner wind region, viz.
\begin{equation}
L_X=\int_{R_{rs}}^{R_c}\Lambda\frac{\rho_{w,d}^2(r)}
{m_{\mu}^2}4\pi r^2dr\ ,
\end{equation}
where $m_{\mu}$ is the proportion of gas mass density to proton
number density. For a H II region, we simply have $m_{\mu}=m_p$,
while $m_{\mu}=1.18m_p$
for gas with a helium abundance of He$/$H$=0.1/0.9$ by number.
Defining the X-ray parameter $X\equiv\Lambda\beta^2/m_{\mu}^2$,
we readily obtain the following expression
\begin{equation}\label{xrayluminos}
L_X=4\pi X\big({1}/{R_{rs}}-{1}/{R_c}\big)\ .
\end{equation}
A normalized X-ray surface brightness profile ${\cal B}$ can be
calculated by integrating over the X-ray luminosity along the line
of sight. Let $R$ be the projected radius from the centre of a
grossly spherical PN, we have the X-ray surface brightness radial
profile ${\cal B}$ as
\begin{equation}
\frac{\cal B}{2}=\left\{\begin{aligned}
\int_{\sqrt{R_{rs}^2-R^2}}^{\sqrt{R_c^2-R^2}}
\Lambda\frac{\rho_{w,d}^2(\sqrt{R^2+x^2},t)}
{m_{\mu}^2}dx\ ,\ \ 0\leq R<R_{rs}\\
\int_0^{\sqrt{R_c^2-R^2}}\Lambda
\frac{\rho_{w,d}^2(\sqrt{R^2+x^2},t)}{m_{\mu}^2}dx\ ,
R_{rs}\leq R<R_c\\
0\ .\qquad\qquad\qquad\qquad R>R_c\\
\end{aligned}
\right.
\end{equation}
By introducing a dimensionless function $L(z)$ defined as
\begin{equation}
L(z)=\frac{1}{z^3}\left[z(1-z^2)^{1/2}+\arccos{z}\right]\,  \
\hbox{ for }\ 0<z<1\ ,
\end{equation}
we immediately arrive at the following expression
\begin{equation}\label{xraybrightness}
\!\!\!\frac{{\cal B}(R,t)}{X/R_c^3}=\left\{\begin{aligned}
\!\!\!\!\!\! \quad L\left(\frac{R}{R_c}\right)
-\frac{R_c^3}{R_{rs}^3}L\left(\frac{R}{R_{rs}}\right)\ ,
\ 0\leq R<R_{rs}\ \ \ \qquad\ \ \ \ \\
\!\!\!\!\!\!\!\!\!\!\!\!\!\!\!\!\!\!\!\!\!\!\!
L\left(\frac{R}{R_c}\right)\ ,
\qquad\qquad R_{rs}\leq R<R_c\qquad\qquad \ \  \\
0\ ,\qquad\qquad\quad
\qquad\qquad\quad R>R_c\ .\ \qquad\qquad \\
\end{aligned}
\right.
\end{equation}
Figures \ref{fig:XrayBrightness} and \ref{fig:XrayMap} illustrate
several model PN X-ray brightness ${\cal B}$ profiles marked with
different values of radius ratio $R_{rs}/R_c$ within the range for
this normalized reverse shock radius $0<R_{rs}/R_c<1$. We
emphasize that morphologies of such diffuse X-ray emissions from
PNe are determined by this radius ratio $R_{rs}/R_c$. The case of
$R_{rs}/R_c$ close to unity corresponds to a ring-like brightness
profile (e.g. the case of PN NGC 40), while a small $R_{rs}/R_c$
gives a more spherical appearance for the X-ray surface brightness
radial profile with a more luminous central bulge or sphere (e.g.
the case of PN BD +30$^{\circ}$3639).
Both X-ray luminosity and surface brightness radial profile depend
only on the X-ray parameter $X$ and radius ratio for reverse shock
and contact discontinuity. The X-ray parameter $X$ is a function
of $\Lambda$ (or inner shocked downstream wind plasma temperature
$T_{w,d}$), $\beta$ and $m_{\mu}$ while parameter $\beta$ is
further determined by the mass loss rate $\dot{M}_{fw}$ from the
central white dwarf, reverse shock speed $v_{rs}$, inner fast wind
velocity $v_{w,u}$ and shocked downstream wind velocity $v_{w,d}$
according to expression (\ref{shocked_fastwind_density}). As the
inner radius of the dense shell, the contact discontinuity radius
$R_c$ can be measured using observations in optical and infrared
bands (e.g. Kastner et al 2000; Guerrero et al 2004; Montez et al.
2005). Parameters $X$ and $R_{rs}$ can be inferred by fitting an
X-ray surface brightness radial profile and the estimate of total
X-ray luminosity of a PN. Spectral analysis can measure the
innermost fast wind velocity, dense shell gas
velocity\footnote{Relation (\ref{CD_requirement}) indicates that
the dense shell gas velocity also gives the inner shocked
downstream wind velocity as required by the pressure balance
across an expanding contact discontinuity radius $R_c$.} and
temperatures (four in all) in the inner and outer zones (e.g.
Guerrero et al 2004; Montez et al. 2005). Pressure balance
condition (\ref{CD_balance}) across the expanding contact
discontinuity radius $R_c$ further connects physical parameters in
the inner shocked downstream wind region and the dense shell. We
emphasize that combined observations in infrared, optical and
X-ray bands may be utilized to infer all relevant parameters that
characterize a quasi-spherical PN.

The morphology of diffuse X-ray emission from a PN only depends on
the radius ratio $R_{rs}/R_c$ (see Figs. 1 and 2). Thus the time
evolution of this radius ratio determines how the X-ray morphology
evolves with time $t$. In other words, different epochs of PN
evolution may be invoked to explain why some PNe are observed to
be strong X-ray sources while others are not. Here, we present
straightforward calculations to show the relation between the
physical evolution state of PNe with their diffuse X-ray emission
morphologies. Suppose the contact discontinuity surface has a
radius $R_{c0}$ at the time when the central fast wind collides
with the slow dense AGB wind envelope. Let $t^{*}$ be the time
lapse from when the time
forward and reverse shock pair initiated after
a wind-wind collision to the current epoch. We
may use $R_{c0}$ and $v_c$ to estimate a kinematic timescale
$t\sim R_{c0}/v_c +t^{*}$ and regard this timescale $t$ as the
dynamic age of a PN (see self-similar transformation equation
\ref{ST} above). It follows that $(t-t^*)$ roughly represents the
temporal duration that the central star has depleted its hydrogen
envelope but not yet started to launch a fast hot tenuous wind.
Then the radius ratio $R_{rs}/R_c$ evolves with time as
\begin{equation}\label{radiusratio}
\frac{R_{rs}}{R_c}=\frac{(R_{c0}+v_{rs}t^*)}{(R_{c0}+v_ct^*)}
=\frac{v_ct-(v_c-v_{rs})t^*}{v_ct}\ .
\end{equation}
At the moment that wind-wind collision occurs and both forward and
reverse shocks are initiated, the radius ratio $R_{rs}/R_c=1$ and
no diffuse X-ray is emitted (see eqn. \ref{xrayluminos}). If PNe
are identified around this stage, then there would be no or very
weak X-ray emissions. The time evolution of $R_{rs}/R_c$ is mainly
determined by $v_{rs}$ according to expression
(\ref{radiusratio}). If $v_{rs}$ is nonnegative, then this radius
ratio has a limiting value of $v_{rs}/v_c$ in the long term
evolution for a large $t^{*}$ (see Fig. 3) during which the
initial contact discontinuity radius $R_{c0}$ becomes less
important. Otherwise, when the reverse shock travels radially
inwards, the region inside the contact discontinuity surface tends
to be filled with more shock heated downstream wind plasma.
Meanwhile, gravity, magnetic field and radiation of the central
star at the shrinking radius of the reverse shock will eventually
become sufficiently strong, and neither asymptotic central wind
profile nor shock condition (\ref{general_shock}) is reliable any
more.
We will show presently that the energy conservation actually
forbids $R_{rs}\rightarrow 0$.


We now consider this problem from the perspective of energy
conservation. The inner fast wind kinetic energy and thermal
energy input per unit time
\begin{equation}
\frac{dE_{in}}{dt}=\frac{1}{2}\dot{M}_{fw}v_{w,u}^2
+\frac{3}{2}\dot{M}_{fw}a_{w,u}^2
\end{equation}
should be larger than the kinetic and internal energy increase of
the system when X-ray is emitted. The internal energy change rates
of fast wind zone and shocked fast wind zone are
\begin{multline}
\frac{dU_{w,u}}{dt}=\frac{3}{2}4\pi R_{rs}^2 v_{rs}
\rho_{w,u}(R_{rs})a_{w,u}^2=\frac{3}{2}\frac{v_{rs}}
{v_{w,u}}\dot{M}_{fw}a_{w,u}^2\ ,\\
\frac{dU_{w,d}}{dt}=\frac{3}{2}4\pi R_{c}^2 v_{c}
\rho_{w,d}(R_{c})a_{w,d}^2-\frac{3}{2}4\pi R_{rs}^2 v_{rs}
\rho_{w,d}(R_{rs})a_{w,d}^2\\
=\frac{3}{2}(1-v_{rs}/v_{w,u})\dot{M}_{fw}a_{w,d}^2\ ,\quad\quad
\quad\quad\quad\quad\quad\quad
\end{multline}
respectively. The kinetic energy change rate caused by the reverse
shock in the inner wind is
\begin{multline}
\frac{dK}{dt}=4\pi
R_{rs}^2\bigg[(v_{w,d}-v_{rs})\frac{\rho_{w,d}v_{w,d}^2}{2}
-(v_{w,u}-v_{rs})\frac{\rho_{w,u}v_{w,u}^2}{2}\bigg]\\
=\frac{\dot{M}_{fw}}{2}(1-v_{rs}/v_{w,u})(v_{w,d}^2-v_{w,u}^2)<0\
.\quad\quad\quad
\end{multline}
The power of the shocked downstream wind zone (hot bubble) pushing
the dense shell outwards is
\begin{equation}
\frac{dW}{dt}=4\pi R_c^2v_{w,d}\rho_{w,d}a_{w,d}^2
=\dot{M}_{fw}a_{w,d}^2
\frac{(1-v_{rs}/v_{w,u})}{(1-v_{rs}/v_{w,d})}\
.
\end{equation}

Given typically estimated parameter values of a PN $v_{w,u}\approx
10^3$ km s$^{-1}$, $v_{w,d}=v_c\approx10^{1\sim 2}$ km s$^{-1}$,
$v_{rs}\approx10^{0\sim1}$ km s$^{-1}$, $a_{w,u}\approx10$ km
s$^{-1}$ and $a_{w,d}\approx10^{1\sim2}$ km s$^{-1}$ as inferred by
\citeauthor{LZ}, we find that
\begin{multline}
\frac{dE_{in}}{dt}\approx10^{2\sim3}\frac{dU_{w,d}}{dt}\ ,\qquad
\frac{dU_{w,d}}{dt}\approx10^{3}\frac{dU_{w,u}}{dt}\ ,\\
\frac{dE_{in}}{dt}\approx-\frac{dK}{dt}\ ,\qquad\quad\
\frac{dE_{in}}{dt}\approx10^{1\sim3}\frac{dW}{dt}\ .\qquad\qquad
\end{multline}
The energy conservation then requires that
\begin{multline}\label{EnergyConserve}
\!\!\!\!\!\! L_X\leq\frac{dE_{in}}{dt}-\frac{d}{dt}\left(U_{w,u}
+U_{w,d}+K+W\right)\approx\dot{M}_{fw}v_{w,u}^2 .\!\!\!
\end{multline}

Energy conservation requirement (\ref{EnergyConserve}) must be
necessarily met. Using expression (\ref{xrayluminos}), we find
that equation (\ref{EnergyConserve}) limits the possible range of
the hot bubble temperature $T_{w,d}$ and the reverse shock radius
$R_{rs}$. As the X-ray luminosity $L_x$ approaches the upper limit
$\dot{M}_{fw}v_{w,u}^2$, either the cooling becomes dominantly
important so that the temperature $T_{w,d}$ drops rapidly, or the
reverse shock radius $R_{rs}$ can no longer shrink. However, in
reality, it is unlikely that X-ray emissions from the hot bubble
reach the upper limit because cooling effects would become
extremely important before $L_X\sim\dot{M}_{fw}v_{w,u}^2$.
In the above discussion, we have ignored the gravitational
potential energy change rate of the shocked downstream zone
occupied by the hot bubble.

\begin{figure}
\centering
\includegraphics[width=0.50\textwidth]{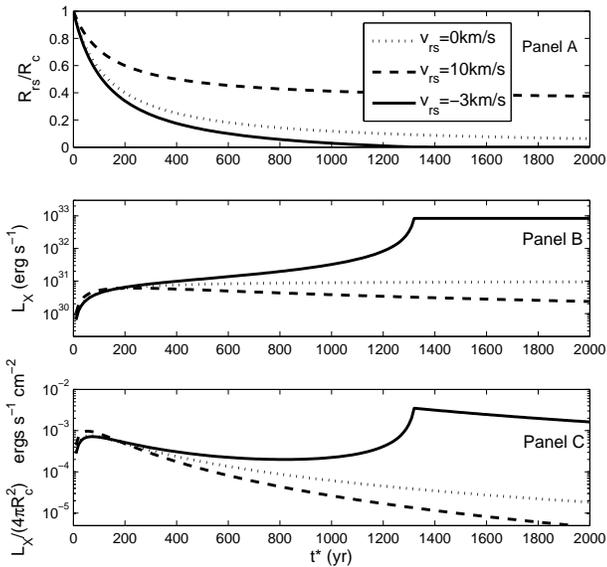}
\caption{
Temporal evolutions of the radius ratio $R_{rs}/R_c$ (Panel A),
the total X-ray luminosity (Panel B) and the average X-ray surface
brightness (Panel C) of three illustrating tracks for PNe with
different reverse shock velocities $v_{rs}$. The solid, dotted and
dash curves in three panels show how PNe evolve when $v_{rs}=-3$,
$0$ and $10$ km s$^{-1}$, respectively. All three models have the
same innermost fast wind mass loss rate
$\dot{M}_{fw}=10^{-8}$M$_{\odot}$\ yr$^{-1}$, contact
discontinuity radius expansion speed $\dot R_c=v_c=30$ km
s$^{-1}$, inner shocked downstream wind plasma X-ray emissivity
$\Lambda=10^{-22}$ ergs cm$^3$ s$^{-1}$ and $R_{c0}=1.3\times
10^{16}$ cm (the distance that a wind blows by at $20$ km s$^{-1}$
for $\sim 200$ yrs).
An inner fast wind velocity of $\sim
500$ km s$^{-1}$ is adopted here.
}\label{fig:RatioEvolution}
\end{figure}

Fig. \ref{fig:RatioEvolution} illustrates how the radius ratio
$R_{rs}/R_c$, the total X-ray luminosity and the average surface
brightness of PNe evolve with negative, zero and positive reverse
shock velocities $v_{rs}$. While all three conditions give
decreasing radius ratio $R_{rs}/R_c$ in general, the temporal
evolution of total X-ray luminosity and average surface brightness
differ from each other significantly.
In the case of $v_{rs}=-3$ km $s^{-1}$, the reverse shock radius
$R_{rs}$ stops at $\sim 10$ au as required by energy conservation
(\ref{EnergyConserve}). We will show presently that for both PNe
NGC 40 and NGC 7662, X-ray emissions only take up $\lsim 1\%$ of
the inner fast wind energy input. Calculations show $R_{rs}\approx
1000$ au when $L_X$ reaches $\sim 1\%$ of $\dot{M}_{fw}v_{w,u}^2$,
viz. $L_X=10^{31}$ erg s$^{-1}$. This appears consistent with the
earlier estimate of Frankowski \& Soker (2009).

\section{Observations of sample PNe}
\subsection{Planetary Nebula NGC 40}

NGC 40 (PNG120.0+09.8; \citealt{Ackeretal1992}) is a very low excitation
PN powered by a WC8 star evolving towards a white dwarf. Within a
ring of nebulosity revealed by optical and near-infrared images of
NGC 40, Montez et al. (2005) detected faint, diffuse X-ray
emission distributed as a partial annulus using the data from the
\emph{Chandra X-Ray Observatory} (ObsID 4480). This annulus-like
X-ray profile corresponds to a fairly large ratio of $R_{rs}/R_c$
and thus a low X-ray luminosity (see equations \ref{xrayluminos}
and \ref{xraybrightness} and Figs. \ref{fig:XrayBrightness} and
\ref{fig:XrayMap}). The actual inferences are X-ray temperature
$T=(8.0\pm 2.0)\times 10^5$ K and total luminosity $L_X\sim
1.5\times 10^{30}\ (D/1.0$kpc$)^2$ ergs s$^{-1}$ of NGC 40 (Montez
et al. 2005).
Kastner et al. (2008) reprocessed the X-ray
data of NGC 40 and updated the luminosity value to $L_X\sim
4.0\times 10^{31}$erg s$^{-1}$ which is adopted in our model
calculations. The X-ray temperature and luminosity are one of
the lowest measured so far for PNe with diffuse X-ray emissions.
Here, the distance to NGC 40 $D=1.0$ kpc is estimated by
Leuenhagen et al. (1996).

\begin{figure}
\centering
\includegraphics[width=0.50\textwidth]{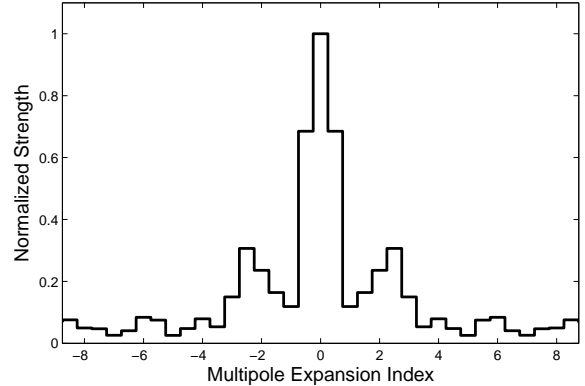}
\caption{Multipole expansion to X-ray counts distribution of NGC
40. Equivalently, this multipole spectrum is the Fourier
transformation of the polar angle distribution of X-ray brightness
map into $\phi-$space, where $\phi\in[-\pi, +\pi]$. The $x-$axis
is the index of multipole and the $y-$axis is the normalized
absolute value of multipole spectrum so that the central strength
is $1$.} \label{fig:NGC40Multipole}
\end{figure}

We decompose the X-ray counts distribution in terms of
cylindrical harmonics in polar coordinates in the plane of sky and
demonstrate the resulting multipole spectrum in Figure
\ref{fig:NGC40Multipole} above. This spectrum reveals a strong
polar uniform component (multipole index $=0$) and a distinct
dipole component (multipole index $=2$) which is about one third
times as strong as the uniform component. While being somewhat
crude, this suggests an exploration for the X-ray brightness
radial profile of NGC 40.

\begin{figure}
\centering
\includegraphics[width=0.50\textwidth]{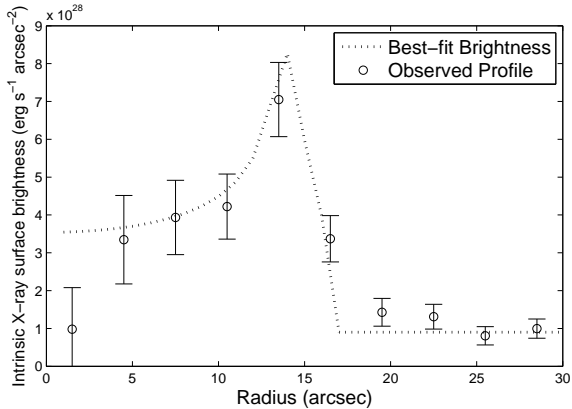}
\caption{Observed X-ray surface brightness profile (circles with error bars
and our model best-fit curve (dotted curve) for the PN NGC 40 are
shown here. The data was taken from \emph{Chandra X-Ray
Observatory} (ObsID 4480) in Montez et al. (2005). The radial
profile was constructed by (i) regarding the X-ray centroid as the
centre of NGC 40, (ii) summing X-ray photon counts in each of a
specified set of concentric circular annuli, (iii) taking the
error in each annulus for Poisson statistics (square root of the
number of photon counts), and (iv) dividing the counts and errors
by the area of each annulus bin. The bin width for each annulus is
chosen as $\sim 3''$. A uniform background X-ray flux $\sim
3.4\times 10^{26}$ ergs s$^{-1}$ arcsec$^{-2}$ is estimated in the
circular region of radius $\gsim 25''$.
}\label{fig:NGC40Fit}
\end{figure}

\begin{figure}
\centering
\includegraphics[width=0.5\textwidth]{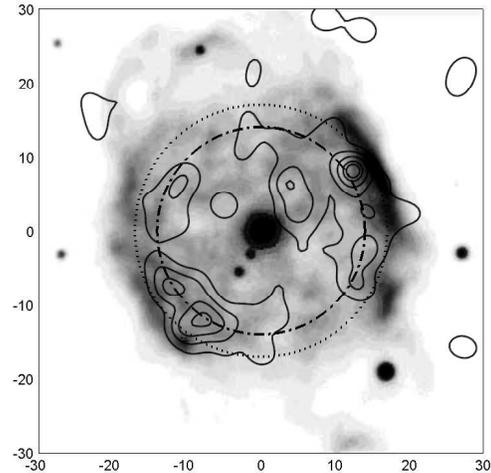}
\caption{
Contours of the smoothed Chandra X-ray image overlaid on
the optical BVR
composite gray-scale image of NGC 40 (Montez et al. 2005) with
model-fitted dash-dotted $R_{rs}$ circle (reverse shock) and
dotted $R_c$ circle (contact discontinuity; $R_c>R_{rs}$).
The X-ray and optical images are obtained
by Chandra and WIYN observatory, respectively. The WIYN
Observatory is a joint facility of the University of
Wisconsin-Madison, Indiana University, Yale University, and the
National Optical Astronomy Observatories.
}
\label{fig:NGC40_overlap}
\end{figure}

The observed X-ray brightness radial profile of NGC 40 is
displayed in Fig. \ref{fig:NGC40Fit} (i.e.the open circles with
error bars). We assume a uniform X-ray emission background
approximately and fit our PN model results (i.e. the dotted curve
in Fig. \ref{fig:NGC40Fit}) with the observed data profile. The
best-fit model gives a radius ratio of $R_{rs}/R_c\sim 0.8$
and an X-ray parameter $X=
2.83\times 10^{48}$ $(D/1.0$ kpc$)^{3}$ ergs s$^{-1}$ cm.
By equation (\ref{xrayluminos}), these results further lead to an
X-ray luminosity from NGC 40
as $L_X\sim 3.7\times 10^{31}(D/1.0$ kpc$)^{2}$ ergs s$^{-1}$,
consistent very well with the observational inference (e.g.
Kastner et al. 2008).
The model fit also gives a reverse shock radius $R_{rs}\approx
14''D=2.0\times 10^{17}(D/1.0$ kpc$)$ cm and a contact discontinuity
radius $R_c\approx 17''D=2.5\times 10^{17}(D/1.0$ kpc$)$ cm,
corresponding to the peak and boundary of diffuse X-ray emissions
from shocked hot bubble. In Figure \ref{fig:NGC40_overlap}, we
circle these model-fitted $R_{rs}$ and $R_c$ on the optical and
X-ray composite image of PN NGC 40. By visual inspection, $R_{rs}$
does pass through the X-ray brightest region and $R_{c}$ roughly
marks the boundary of the diffuse X-ray emission. \citeauthor{LZ}
piecewise isothermal model claims that the optically bright PN rim
is located just outside the contact discontinuity radius $R_c$ and
the density becomes highest near the forward shock radius $R_s$.
Figure \ref{fig:NGC40_overlap} indicates that the forward
shock is just outside the contact discontinuity surface so that
the optical bright dense shell is very thin, i.e. $R_s\gtrsim
R_c\sim 17''$.
According to the spectral analysis of Sabbadin et
al. (2000), the most optically luminous shell region of NGC 40
still have redshift and blueshift componeonts at radius $\gsim
17''$, implying $R_s\gsim 17''$. Figure \ref{fig:NGC40_overlap}
shows that our model fits observation very well.
The radial separation between the most optically bright radius at
$\sim 17''$ and the most X-ray bright radius at $\sim 14''$ is
clearly distinguishable, strongly supporting \citeauthor{LZ}
piecewise isothermal model. In contrast, the adiabatic model of
Chevalier \& Imamura (1983) would claim no significant separation
between these two radii (e.g. Akashi et al. 2006).

The best-fit Raymond-Smith thermal plasma emission model
(Raymond \& Smith 1977)
with the X-ray spectrum of NGC 40 indicates a plasma temperature
of $T_X=(8.0\pm 2.0)\times 10^5$ K (Montez et al. 2005); at such
temperature, the spectral line cooling from highly ionized metals
is important
while the thermal bremsstrahlung emission from free hot electrons
is relatively weak,
and the total X-ray emissivity is
$\Lambda\approx 2.0\times10^{-22}$ ergs cm$^3$ s$^{-1}$
(e.g., Sarazin 1986). With $m_{\mu}=m_p$, we have
$\beta=3.16\times 10^{-8}$M$_{\odot}$yr$^{-1}/(100$km s$^{-1})$
by expression (\ref{shocked_fastwind_density}). We adopt
$v_{w,d}=v_c=30$ km s$^{-1}$ as
estimated by optical spectrum line analysis near the contact
discontinuity surface (e.g., Sabbadin et al. 1999). Ignoring the
duration that the central star stopped to blow slow AGB wind and
not yet began to launch tenuous fast wind, we estimate a wind-wind
dynamic interaction timescale of $R_c/v_c\sim 3000$ yrs (a
comparable timescale of $\sim 5000$ yrs was estimated by Akashi et
al. 2006). For the radius ratio
$R_{rs}/R_c=0.8$, we have
$v_c-v_{rs}=6.0$ km s$^{-1}$. Thus the reverse shock moves
outwards in the laboratory framework of reference. Meanwhile,
equation (2) gives an inner tenuous fast wind speed
$v_{w,u}=1.1\times 10^3$ km s$^{-1}$.
Equation (\ref{shocked_fastwind_density}) gives a central star
mass loss rate as
$\dot{M}_{fw}\approx 2.4\times 10^{-8}$ M$_{\odot
}$yr$^{-1}$, which
is very close to an earlier estimate of
$\sim 3\times 10^{-8}$M$_{\odot }\ $yr$^{-1}$ by UV spectral
analysis, and falls within the estimated range of $\sim 10^{-10}-
10^{-7}$M$_{\odot }\ $yr$^{-1}$ (e.g., Cerruti-Sola \& Perinotto
1985). By imposing condition (\ref{CD_balance}), the diffuse X-ray
emission from the central zone of PN NGC 40 requires a
$T_d=1.5\times10^4(10^{-6}/\alpha_0)$ K for the outer shocked
dense shell.
We note that the X-ray emission only takes up $\sim 0.2\%$
of $\dot{M}_{fw}v_{w,u}^2$ for PN NGC 40.

However, our best-fit model in Fig. \ref{fig:NGC40Fit}
overestimates the X-ray brightness
for smaller radii around the center. In \citeauthor{LZ} model, the
temperature of the dense shell region is not high enough to radiate
in X-ray bands. In reality, thermal conduction inevitably occurs
across the contact discontinuity so that the dense shell region will
be heated up by inner shocked downstream wind and may lead to
partial X-ray
emissions.
The inner shocked downstream wind zone near the
contact discontinuity surface would also become denser and cooler
due to diffused materials and thermal conduction from the dense
shell, especially if magnetic field is absent (e.g., Steffen et
al. 2008). This effect would lead to a relatively higher X-ray
flux than our model result around the outer boundary of the hot
bubble.
Nevertheless, this thermal conduction is expected to be sensitively
suppressed even for the presence of a weak magnetic field transverse
to the radial direction (e.g., Chevalier 1997; Yu et al. 2006; Wang
\& Lou 2008; Lou \& Hu 2009; \citeauthor{LZ}), so the total X-ray
luminosity of PNe is more or less confined to emissions within
radius $R_c$. Within the radius of $\sim 2$ arcsec or $\sim
3.0\times 10^{16}$ cm or $\sim 2000$ au around the central region,
X-ray brightness of our model is $\sim 75\%$ greater than the upper
limit of confidence interval of observation (see Fig.
\ref{fig:NGC40Fit}).
Since the reverse shock is at $\sim
14''$ within which there is no diffused X-ray emission, the
projected X-ray photons within radius $\sim 2''$ are actually
emitted in the hot bubble at larger radii closed to $R_c$ but
projected around the center along the line of sight. If there is a
dense dust cloud surrounding the central star, the X-ray photons
emitted behind the cloud might be partially absorbed so that the
observed central X-ray brightness becomes lower than expected.
Both theoretical calculations (e.g., Frankowski \& Soker 2009) and
observations show that dense gas/cool dust clouds may exist around
the central white dwarf of PNe with diameter $\gsim 150$ au (e.g.
Helix nebula, dust density $\sim 2.5$ g cm$^{-3}$ by Su et al.
2007). The cross section of such a dust cloud is about $2\times
10^4$ au$^2$, which can only obscure $\sim 0.5\%$ X-ray emission
within $\sim 2000$ au radius maximally. If this is the case for PN
NGC 40, Figure \ref{fig:NGC40Fit} requires a column electron
number density of $\sim 2.7\times 10^{24}$ cm$^{-2}$ within $\sim
2''$ to absorb $\sim 80\%$ X-ray emissions behind central star by
Thomson scattering. Such a high column density corresponds to a
total cloud mass $\sim 5M_{\odot}$. We thus conclude that X-ray
absorption by either dust or gas cloud cannot explain the
unusually low X-ray brightness in the central NGC 40.
Alternatively, the non-spherical structure of NGC 40 might be the
most important factor.


The dipolar structure of the X-ray brightness is noticeable in PN
NGC 40. Also, the eastern X-ray region is brighter and seems to be
more confined to within the brighter optical rim. As a first
approximation, our spherically symmetric treatment modelling here
ignore these asymmetries.

\subsection{Planetary Nebula NGC 7662}


NGC 7662 appears to be a quasi-spherical PN with a moderately large
diameter of $\sim 30''$ or $\sim 4\times 10^{17}$ cm and with
optical surface brightness but without X-ray data available up to
the present. Guerrero et al. (2004) probed the structure and
kinematics of NGC 7662 based on long-slit echelle spectroscopic
observations and on Hubble Space Telescope (HST) archival
narrow-band images. They inferred that PN 7662 consists of a central
cavity surrounded by two concentric shells, viz. the shocked dense
shell and the outer slow AGB wind envelope, and estimated the
temperatures of the two shells as $\sim 1.4\times 10^4$ K and $\sim
1.1\times 10^4$ K, respectively. By fitting the radial density
profiles of the dense shell zone (see figure 13 of \citeauthor{LZ})
and the slow AGB wind envelope with the self-similar dynamic
evolution model of \citeauthor{LZ} for isothermal voids,
\citeauthor{LZ} estimated relevant physical parameters for PN NGC
7662 which appear consistent with optical observations (e.g.,
Guerrero et al. 2004). \citeauthor{LZ} have also inferred properties
of the inner fast wind by imposing condition (\ref{CD_balance}) for
the pressure balance across the contact discontinuity. Shocked
downstream inner wind velocity $v_{w,d}=25.9$ km s$^{-1}$ was
inferred by the outer dense shell expansion. The innermost fast wind
mass loss rate $\dot{M}_{fw}\approx 2\times 10^{-8}M_{\odot}$
yr$^{-1}$, the inner fast wind speed $v_{w,u}=1500$ km s$^{-1}$, the
hot bubble (i.e. shocked downstream inner wind zone) plasma
temperature $\sim 6.4\times 10^6$ K and the reverse shock velocity
$v_{rs}=-10$ km s$^{-1}$ represent a plausible set of fitting
parameters that can reproduce a PN with more or less the same outer
zone of NGC 7662. At the current stage of evolution, we have
estimated the two radii $R_c=5.1\times 10^{16}$ cm and
$R_{rs}=6.5\times 10^{15}$ cm, i.e. the radius ratio
$R_{rs}/R_c=0.127$ is fairly small. Equation
(\ref{shocked_fastwind_density}) gives the $\beta$ value as $\sim
4.43\times 10^{-11} $M$_{\odot }\ $yr$^{-1}/($km s$^{-1})$. The
total X-ray emissivity is $\Lambda\approx 5.1\times 10^{-23}$ ergs
cm$^3$ s$^{-1}$ for a plasma of temperature $6.4\times 10^6$ K
(e.g., Sarazin 1986). For $m_{\mu}=m_p$, we then have an X-ray
parameter $X=1.42\times 10^{46}$ ergs s$^{-1}$ cm. This would give a
total X-ray luminosity of NGC 7662 as $L_X=2.4\times 10^{31}$ ergs
s$^{-1}$.
In short, we predict that an X-ray surface brightness like the
upper-left plot of Fig. \ref{fig:XrayBrightness} with no distinct
reverse shock surface and a total X-ray luminosity similar to that
of NGC 40 is anticipated for NGC 7662, which should be detectable
by the \emph{Chandra X-Ray Observatory} in space.

\begin{figure}
\centering
\includegraphics[width=0.50\textwidth]{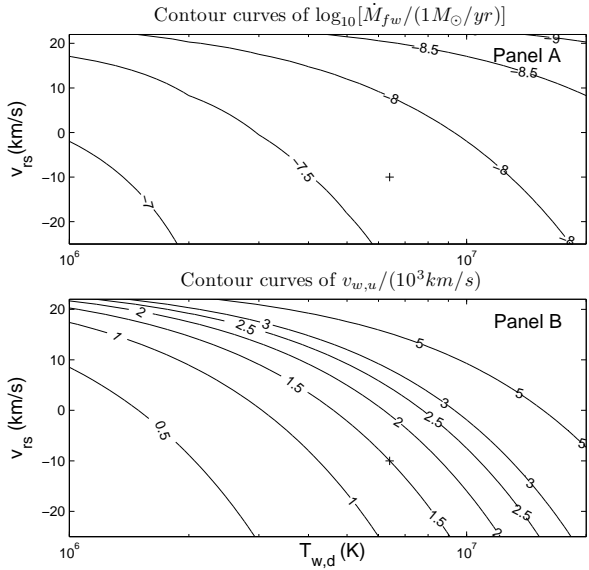}
\caption{
Piecewise isothermal model solution properties of inner wind zone
are explored for the PN NGC 7662. Panel A shows a chosen set of
contour curves for $\log_{10}\dot{M}_{fw}$ in the unit of
$M_{\odot}$ yr$^{-1}$ in a $v_{rs}$ versus $T_{w,d}$ presentation.
Panel B shows a chosen set of contour curves for the inner fast
wind speed $v_{w,u}$ in the unit of $1000$ km s$^{-1}$ in a
$v_{rs}$ versus $T_{w,d}$ presentation. The $T_{w,d}$ axes of both
panels are in the logarithmic scale. As discussed in the main
text, the example model solution with parameters $v_{rs}=-10$ km
s$^{-1}$, $T_{w,d}=6.4\times 10^6$ K, $\dot{M}_{fw}=2\times
10^{-8} M_{\odot}$ yr$^{-1}$, $v_{w,u}=1500$ km s$^{-1}$ is marked
here by the plus symbol $+$.}\label{fig:SolutionNGC7662}
\end{figure}

\begin{figure}
\centering
\includegraphics[width=0.50\textwidth]{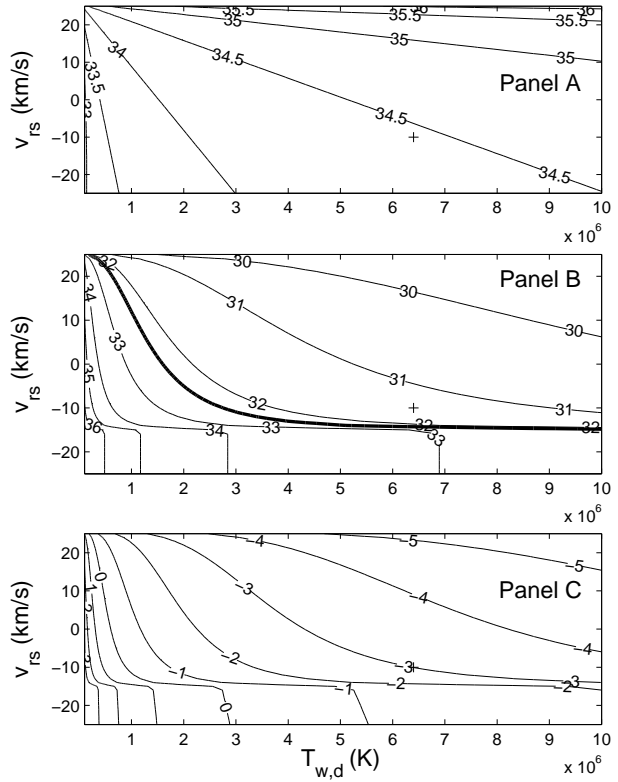}
\caption{
Energy solutions of inner wind zone properties for the PN NGC
7662. Following Figure \ref{fig:SolutionNGC7662}, here we show:
(Panel A) a chosen set of contour curves for
$\log_{10}(\dot{M}_{fw}v_{w,u}^2)$ in the unit of erg s$^{-1}$ in
a $v_{rs}$ versus $T_{w,d}$ presentation; (Panel B) a chosen set
of contour curves for $\log_{10}L_X$ in the unit of erg s$^{-1}$
in a $v_{rs}$ versus $T_{w,d}$ presentation; (Panel C) a chosen
set of contour curves for $\log_{10}[L_X/(\dot{M}_{fw}v_{w,u}^2)]$
in a $v_{rs}$ versus $T_{w,d}$ presentation. For computing $L_X$,
a minimum $R_{rs}=10$ AU is adopted. The heavy solid curve in
Panel B marks the upper limit given by ROSAT observation. The
bottom left region to the heavy solid curve gives X$-$ray
luminosity higher than the upper limit, which is forbidden from
the energetic consideration. The special model solution
$v_{rs}=-10$ km s$^{-1}$, $T_{w,d}=6.4\times 10^6$ K,
$\dot{M}_{fw}=2\times 10^{-8} M_{\odot}$ yr$^{-1}$, $v_{w,u}=1500$
km s$^{-1}$ is here marked by the plus symbol $+$.
}\label{fig:EnergyNGC7662}
\end{figure}

However without available X-ray observations of NGC 7662, physical
conditions of its inner wind zone confined within the contact
discontinuity radius $R_c$ of the PN NGC 7662 cannot be uniquely
determined at this stage of model analysis. In other words, in
terms of fitting available observations of NGC 7662 in other bands
(e.g. optical, infrared bands and so forth), we can predictively
explore X-ray emissions from the inner shocked downstream wind
plasma of NGC 7662. In the above procedure, the innermost fast
wind mass loss rate is a key yet adjustable parameter and may vary
within a sensible range. We now show the corresponding ranges of
our model results for NGC 7662 in Figs. \ref{fig:SolutionNGC7662}
and \ref{fig:EnergyNGC7662}.
In fact, the \citeauthor{LZ} ISSV model solution with a parameter
set $\{ v_{rs}=-10$ km s$^{-1}$, $T_{w,d}=6.4\times 10^6$ K,
$\dot{M}_{fw}=2\times 10^{-8}\ M_{\odot}$ yr$^{-1}$, $v_{w,u}=1500$
km s$^{-1} \}$ is only one of the plausible inner shocked wind
solutions that meet pressure balance condition (\ref{CD_balance})
across the contact discontinuity. By systematically varying the
inner downstream shocked wind temperature $T_{w,d}$ and the reverse
shock speed $v_{rs}$, we readily derive different innermost fast
wind velocities $v_{w,u}$ by equation (\ref{reverse_shock}) and the
mass loss rates $\dot{M}_{fw}$ of the central star by wind pressure
balance condition (\ref{CD_balance}) across the contact
discontinuity as shown in Fig. \ref{fig:SolutionNGC7662}. To be
specific, we pick a solution example marked by the plus symbol $+$
in Figure \ref{fig:SolutionNGC7662}.

Figure \ref{fig:EnergyNGC7662} presents the upper limit of
X-ray luminosity $\sim\dot{M}_{fw}v_{w,u}^2$, the predicted X-ray
luminosity $L_X$ by equation (\ref{xrayluminos}) and the ratio
between these two quantities. The PSPC ROSAT count rate limit is
$\sim 0.03 $ s$^{-1}$. If NGC 7662 is not detected in the ROSAT
All Sky Survey (RASS), then webPIMMS
(http://heasarc.gsfc.nasa.gov/Tools/w3pimms.html) puts the upper
limit for the X-ray luminosity of $\sim 2.2\times 10^{32}$ erg
s$^{-1}$ (a typical hydrogen column number density $n_H\sim
10^{21}$ cm$^{-2}$ and a hot bubble temperature $T_{w,d}=2\times
10^6$ K are assumed for NGC 7662 at a distance of $D=1$ kpc; e.g.
Guerrero et al. 2004). The heavy solid curve in Panel B of Figure
\ref{fig:EnergyNGC7662} marks this upper limit and suggests a
possible regime for both $T_{w,u}$ and $v_{rs}$ at the upper right
side of the curve. In the regime where $T_{w,u}\geq 10^6$ K and
$v_{rs}\geq -10$ km s$^{-1}$, the ratio
$L_X/(\dot{M}_{fw}v_{w,u}^2)$ is less than $\sim 10^{-2}=1\%$.
This estimate is also valid for PN NGC 40. Comparing Figures
\ref{fig:SolutionNGC7662} and \ref{fig:EnergyNGC7662}, we can also
figure out that the mass loss rate
$\dot{M}_{fw}\lsim 10^{-7.5}$ M$_{\odot}$
yr$^{-1}$ and the fast wind velocity
$v_{w,u}\gsim 10^3$ km s$^{-1}$.
These estimated results appear all consistent with previous
observations and theoretical calculations.


\section{Summary and Discussion}

We invoke the recently constructed ISSV hydrodynamic model of
\citeauthor{LZ} to explore the dynamic shock interaction of the
inner isothermal fast wind with the outer AGB slow dense wind and to
calculate diffuse X-ray emissions from grossly spherical PNe for
available observational comparisons. Our piecewise isothermal wind
dynamic interaction model involves a forward shock into the outer
AGB slow wind envelope, a reverse shock into the inner fast tenuous
wind and a contact discontinuity in between these two shocks.
Different from previous dynamic model like Chevalier \& Imamura
(1983), \citeauthor{LZ} model indicates that the X-ray emission
peaks near the reverse shock, instead of the contact discontinuity
surface between the hot bubble and the PN dense shell. Based on
Chevalier \& Imamura model, Akashi et al. (2006) indicates that
diffuse X-ray emissions from hot bubbles should be in the form of
narrow bright rings. Our piecewise isothermal model calculations
show that the morphology of diffuse X-ray emission is determined by
the radius ratio $R_{rs}/R_c$ such that either ring-like or central
bright X-ray morphology can appear. According to our dynamic model
analysis, a relatively small $R_{rs}/R_c$ ratio is suggested for PN
BD +30$^{\circ}$3639, where a centrally bright diffuse X-ray
emission morphology persists (e.g. Leahy et al. 2000; Kastner et al.
2000).
In contrast, an adiabatic evolution
model of PN cannot give a morphology for central bright diffuse
X-ray emissions. We would expect various morphological cases for a
more systematic survey of PNe.

As current X-ray detectors in space have been powerful enough to
observe extended X-ray morphologies of PNe, we offer quantitative
line-of-sight X-ray brightness radial profiles for candidate PN
sources. In our model framework, the X-ray parameter $X$ (defined
in the main text), the reverse shock radius $R_{rs}$ and the
contact discontinuity radius $R_c$ together completely determine
both the X-ray luminosity and the surface brightness profile. In
perspective, X-ray observations together with optical and infrared
images of grossly spherical PNe are expected to give specific
measurements or estimates for these three key parameters.

The PN NGC 40 radiates intensely in UV bands, giving a valuable
observational platform to test and calibrate our X-ray model
results for diagnostics. The results of our piecewise isothermal
dynamic wind interaction shock model are fitted well with
observations except that the central X-ray brightness is somewhat
higher than that actually observed as shown in Fig.
\ref{fig:NGC40Fit}. NGC 40 appears to have a radius ratio
$R_{rs}/R_c$ close to unity with a relatively low temperature of
$\sim 10^6$ K in the inner shocked downstream wind plasma. These
two facts result in a fairly low X-ray luminosity of
$\sim 4.0\times 10^{31}$ ergs s$^{-1}$ from NGC 40 (see Kastner et
al. 2008 for this update). UV observations estimate a fast wind
speed range of $\sim 1800-2370$ km s$^{-1}$ (e.g., Cerruti-Sola \&
Perinotto 1985; Bianchi 1992). This is about twice our estimated
fast wind speed $\sim 1100$ km s$^{-1}$.
We also note that the dipolar structure of NGC 40 significantly
affects the morphology of diffused X-ray emission. The actual hot
bubble of NGC 40 is not very spherical, but in a dumbbell-like
shape. Most X-ray photons are emitted from east and west regions,
while only a few are from north, south or central regions.
Observationally, spectral line analysis to the optical bright rims
show that the dense shell gas expands at $\sim 40-50$ km s$^{-1}$
in the north-south direction and at $\sim 30$ km s$^{-1}$ in the
east-west direction (e.g. Sabbadin et al. 1999). This indicates
that the earlier AGB slow wind is denser in the east-west
direction so that the collision between the inner fast wind and
the slow AGB wind is more violent in the east-west region. This
also explains why the dense shell is brighter in the east and west
regions. By both quantitative calculation on spherical smoothing
and qualitative analysis on dipolar structure, we conclude that
wind-wind shock dynamic interaction appears natural for the hot
bubble formation and diffused X-ray emissions.

Different from the PN NGC 40, the PN NGC 7662 is here inferred to
have an inward reverse shock velocity of $\sim -10$ km s$^{-1}$.
By expression (\ref{xrayluminos}), NGC 7662 is predicted to have
an increase of X-ray luminosity with increasing time $t$ at an
estimated rate of $\sim 0.9\times 10^{28}$erg s$^{-1}$ per year.
For example, its X-ray luminosity will reach $\sim 2.1\times
10^{30}$ ergs s$^{-1}$ 20 years later and $\sim 2.5\times 10^{30}$
ergs s$^{-1}$ 50 years later.
Energy conservation shows a minimum $R_{rs}$ exists so that $L_X$
must be limited by a maximum $\sim\dot{M}_{fw}v_{w,u}^2$. In fact,
studies on NGC 40 and NGC 7662 both lead to the conclusion that
$L_X\lsim 10^{-2}\dot{M}_{fw}v_{w,u}^2$. The sensible explanation
is that cooling effect becomes very significant even when $L_X$ is
still far lower than this upper limit.
%

At the present epoch, it appears justifiable to not include cooling
effects in our piecewise isothermal dynamic model calculations.
Akashi et al. (2006) considered the radiative cooling
semi-quantitatively for computing the X-ray luminosity based on the
self-similar colliding adiabatic wind model of Chevalier \& Imamura
(1983) and showed that the radiative cooling was unimportant during
most time of a PN dynamic evolution. Nevertheless, as radiative
cooling becomes important for the hot bubble evolution, e.g.
$\sim 10^3\ $yrs later for the case
of $v_{rs}=-3\ $km s$^{-1}$ in Figure \ref{fig:RatioEvolution},
it would be inaccurate to compute X-ray emissions either in our
model or in that of Akashi et al. (2006). The X-ray hot bubble
will cool down much faster, the contact discontinuity balance
condition would be broken and the outer dense shell dynamic
behaviour may be also interrupted.
%

Since X-ray luminosity only takes up $\lesssim 1\%$ of
$\dot{M}_{fw}v_{fw}^2$, the energy conservation consideration in
Section 3 implies that a significant amount of energy should have
been converted to other forms during the entire life of PNe.
Otherwise, the temperature of shocked downstream inner wind should
be $\sim 10$ times higher than the observed temperature inferred
from diffuse X-ray emissions.
However, relatively low temperatures of X-ray emitting gas are
generally inferred in PNe at different evolution stages. We note
that the energy conservation calculation in Section 3 is fairly
general, where we focus on the mass conservation $\dot{M}_{fw}=4\pi
r^2\rho_{w,d}v_{w,d}$ and choose typical observed parameter values
without specifically invoking either the isothermal or adiabatic
assumptions. Therefore we suspect that there must be certain long
lasting mechanisms to explain the low temperature of X-ray emitting
gas, and such mechanisms are not quantitatively taken into account
in either \citeauthor{LZ} piecewise isothermal model or the
adiabatic model of Chevalier \& Imamura (1983). Moreover, such
mechanisms influence more than radiative cooling does so that
shocked hot bubbles do not evolve adiabatically, as implied by the
observation of central bright X-ray emission observation in PN BD
+30$^{\circ}$3639. Thermal conduction and material diffusion might
be non-negligible if the magnetic field near the expanding contact
discontinuity surface becomes sufficiently weak. Another possibility
is that the hot bubble largely between the reverse shock and the
contact discontinuity becomes turbulent to partially account for the
missing energy. The thermal conduction between hot bubble and cool
dense shell could be considerable (e.g. Kastner et al. 2000; Steffen
et al. 2008). The rapid variance of inner fast wind mass loss rate
of central stars might also be essential to account for the
imbalance between fast wind kinetic energy input and low temperature
inferred by diffuse X-ray emissions (e.g. Soker \& Kastner 2003;
Akashi et al. 2007). In Sections 2 and 3, we did not consider such
mechanisms; instead, we choose the shocked fast wind temperature
$T_{w,d}$ (or $X$) as an adjustable parameter for fitting X-ray
observations.

In some PNe with detected X-ray emissions, very distinct bipolar
morphology appears, such as HENIZE 3-1475 (e.g. Sahai et al.
2003), NGC 7026 (e.g. Gruendl et al. 2006), MENZEL 3 (e.g. Kastner
et a. 2003), and so forth. Collimated outflows and jets in
magnetized gas media might be expected to account for these types
of diffuse X-ray emissions. Akashi et al. (2008) performed
two-dimensional numerical simulations of jets expanding into the
slow wind of AGB stars. They proposed that this jet-wind or
outflow-wind interaction model might explain such bipolar
morphology of diffuse X-ray emissions.

\section{Acknowledgement}

This research was supported in part by the National Natural
Science Foundation of China (NSFC) grants 10373009 and 10533020 at
Tsinghua University, the SRFDP 20050003088 and 200800030071, the
Yangtze Endowment, the National Scholarship and the National
Undergraduate Innovation Training Project from the Ministry of
Education at Tsinghua University and Tsinghua Centre for
Astrophysics (THCA).
%

\end{document}